# Correlative Microscopy of Morphology and Luminescence of Cu porphyrin aggregates

M Bahrami[1], S Kraft[1], J Becker[1], H Hartmann[1], B Vogler[1], K Wardelmann[1], H Behle[1], JAAW Elemans[2], I Barke[1], and S Speller[1]

[1] Institute of Physics, University of Rostock, Rostock, Germany

[2] Institute for Molecules and Materials, Radboud University Nijmegen, The Netherlands

Email Sylvia.speller@uni-rostock.de

Keywords: porphyrin, molecule aggregate, atomic force microscopy, fluorescence, phosphorescence, correlative microscopy, exciton

**Abstract.** Transfer of energy and information through molecule aggregates requires as one important building block anisotropic, cable-like structures. Knowledge on the spatial correlation of luminescence and morphology represents a prerequisite in the understanding of internal processes and will be important for architecting suitable landscapes. In this context we study the morphology, fluorescence and phosphorescence of molecule aggregate structures on surfaces in a spatially correlative way. We consider as two morphologies, lengthy strands and isotropic islands. It turns out that phosphorescence is quite strong compared to fluorescence and the spatial variation of the observed intensities is largely in line with the amount of dye. However in proportion, the strands exhibit more fluorescence than the isotropic islands suggesting weaker non-radiative channels. The ratio fluorescence to phosphorescence appears to be correlated with the degree of aggregation or internal order. The heights at which luminescence saturates is explained in the context of attenuation and emission multireflection, inside the dye. This is supported by correlative photoemission electron microscopy which is more sensitive to the surface region. The lengthy structures exhibit a pronounced polarization dependence of the luminescence with a relative dichroism up to about 60%, revealing substantial perpendicular orientation preference of the molecules with respect to the substrate and parallel with respect to the strands.

## 1. Introduction

Organic crystals are attractive materials for opto-electronics and as light harvesting systems [1,2]. They show high flexibility in preparation, so that they can allow one to arrange for landscapes which are beneficial for processes such as field enhancement, energy transport, and charge separation. Upon illumination, Frenkel excitons are generated in molecule crystals. Although related to Wannier excitons in conventional semiconductors Frenkel excitons exhibit higher binding energies and accordingly smaller spatial extents.

For enabling long energy transfer paths, the aim is to transport excitons over large distances through the aggregate. Basically, two mechanisms are considered: in Forster transfer excitons hop though the crystal mediated by dipole-dipole interaction, while in Dexter transfer direct exchange of electrons occurs. Though in Dexter type transfer diffusivity is substantially smaller, longer life times may overcompensate this drawback [3]. Note that Forster energy transfer suffers from shorter radiative lifetimes as a result of a larger transition dipole moment. Dexter transfer is expected to be more robust against disorder; however it implies intermolecular wave functions to overlap, usually implying finite electric conductivity. In organic dyes singlet excitons typically come along with Forster transfer mechanism and fluorescence, while for triplet excitons Dexter transfer and phosphorescence are dominating mechanisms. Triplet excitons can be the result of ordinary intersystem crossing, e.g. involving spin-orbit interaction or charge transfer states [4], but may also occur due to singlet exciton fission [5].

Traditionally, predominantly spectroscopic studies have been employed to assess dynamic characteristics of excitons; recently the number of microscopy studies addressing spatial properties of excitons is evolving. An opposite transport anisotropy for singlet and triplet excitons has been observed in tetracene crystals [6]. For diindenoperylene enhanced optical near-fields have been found especially at domain boundaries of the crystal which is interpreted by means of strong coupling of the tip plasmon with the exciton polariton [7]. Apart from intensity variations, polarization and lifetime maps are acquired and may reveal information on the degree of order in the assembly, on the orientation of the molecules in the aggregates, and on the competition of luminescence versus non-radiative de-excitation pathways. In [8] a polarization dichroism of 80% was determined for two distinct morphologies of thiophene derivatives. This was attributed to different levels of energetic disorder in the assemblies. In case of thiacarbocyanine aggregates the excitation polarization was varied, a stronger emission with polarization along the fiber axes was observed, and it was concluded that the long axis of the molecule is aligned along the fiber axes [9]. Just slightly different exciton life times were reported for rod- and sphere shaped porphyrin aggregates[10]. As a first step to answer the question which types of excitations are present and how they depend on structural features and the environment, we investigate morphology, fluorescence, and phosphorescence of porphyrin aggregate structures by correlative microscopy methods.

## 2. Experiment

We prepared different structures of Cu porphyrin aggregates on substrate surfaces, and studied their morphology, fluorescence, and phosphorescence by correlative AFM and optical microscopy. We used Cu porphyrins functionalized with undecyl tails at the four meso-positions, the synthesis has been described earlier [11]. Side groups are very flexible and therefore enhance solubility, including non-aromatic solvents. Alkyl chains are known to facilitate the binding of molecules on graphite surfaces, as has been shown in STM studies of monolayers [12]. If not otherwise stated we prepared 3D strands by drop casting of 1 to $2\times10^{-4}$ molar porphyrin-heptane solution without any post-rinsing. The substrate type is silicon wafer, highly oriented pyrolytic graphite (HOPG), or glass. In the course of evaporation of the solvent characteristic tree-like structures grow on the surface; occasionally shorter, straight structures or largely isotropic islands occur.

## 3. Results

*3.1 Formation of structure types*

The aggregation is governed by the coffee ring effect [13,14];due to stronger curvature a droplet exhibits higher solvent evaporation speed at its rim. Material flows outwards, leading to higher concentration and pronounced aggregation at the border of the droplet. The size of the rings and its overall shape is determined by an interplay between pinning, surface tension, evaporation, and hydrodynamic and vibrational properties of the droplet. The same mechanisms also affect the morphology of the molecule aggregates. In our case, larger dendritic and smaller isotropic structures form (figure 1(c)). In regions with fast, uninterrupted retraction of the dewetting front, isotropically shaped aggregates form (see figure 1(a)). Aggregation is enhanced if during shrinkage of the droplet the dewetting front transiently gets pinned or otherwise slowed down, e.g. by approaching topographic irregularities on the surface, by dust particles or other contaminations (figure 1(b)). Then, tree-like aggregates form. Occasionally, the dewetting front wobbles and tree aggregates can be dislocated (figure 1(c)), particularly the larger ones.

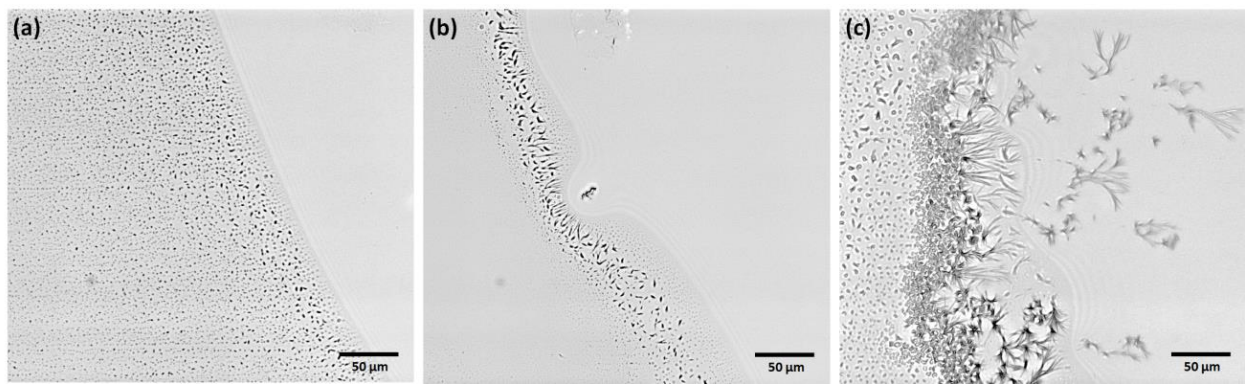

**Figure 1.** Individual frames from a movie (bright field microscopy) during aggregation. The straight borderline corresponds to the dewetting front which is propagating from left to right here. Regions with distinct morphologies, a more isotropic, globular (a) and dendritic shape (b), occur. (c) During retraction of the dewetting front tree-shaped aggregates occasionally become displaced.

*3.2 Morphology, adhesion, and structure*

In the following we describe the morphology of the elongated tree-like and the more isotropic structures. The tree structures are several 10 nm, up to 200 nm high (figure 2(a)) and show fluorescence as well as phosphorescence (figure 2(b),(c), see also section 3.4). The strands have belt-shaped profiles (figure 2(d)) exhibiting a wide spectrum of widths from ~100 nm to ~1000 nm. The lateral morphology is slightly curved. Very striking is the rather uniform splitting angle of around 26°-29° at bifurcations. In certain regions the structures appear to be agglomerates of smaller rod like structures (figure 2(e)). If curved fibers are cleaved by means of the AFM tip, the two ends jump into a straight configuration, revealing that the strands are subject to tension as a consequence of the attachment to the surface (not shown). At the same time this points to rather weak adhesion which is in line with the observation that whole aggregates are dislocated during preparation via forces due to the moving dewetting front (see figure 1(c)). Cutting strands by means of an AFM tip results quite flat planes at the cleft perpendicular to the strand axis; apparently cleavage occurs along a crystallographic plane, which is observed roughly perpendicular to the strand axis. This is a first sign of molecular order in the strands. Below, we show on the basis of polarization dichroism, that the lengthy structures are substantially ordered.

The smallest tree-related structures show heights of ~ 5 nm and lengths of several 100 nm. In contrast to the longer strands these small structures are quite straight suggesting a fully relaxed conformation on the substrate. The larger tree-like aggregates bend on length scales of ~10 µm. For the flatter structures (below ~30 nm) luminescence is below the detection limit of our microscope. The isotropic islands show heights between 70 nm and 120 nm, and diameters in the range of 0.6 µm to 1.5 µm, see figure 3(a),(b). Isotropic structures dominate if the substrate is glass. In case HOPG serves as substrate and the 3D aggregates are removed by rinsing a monolayer of flat-lying molecules exhibiting two characteristic lattices is covering the whole surface [11]. An AFM topography and a phase image of a molecular strand and the domain structure of the monolayer is shown in figure 2(f) and (g), respectively. The adherence of the 3D structures is weak; they are retained on the surface if no rinsing is applied.

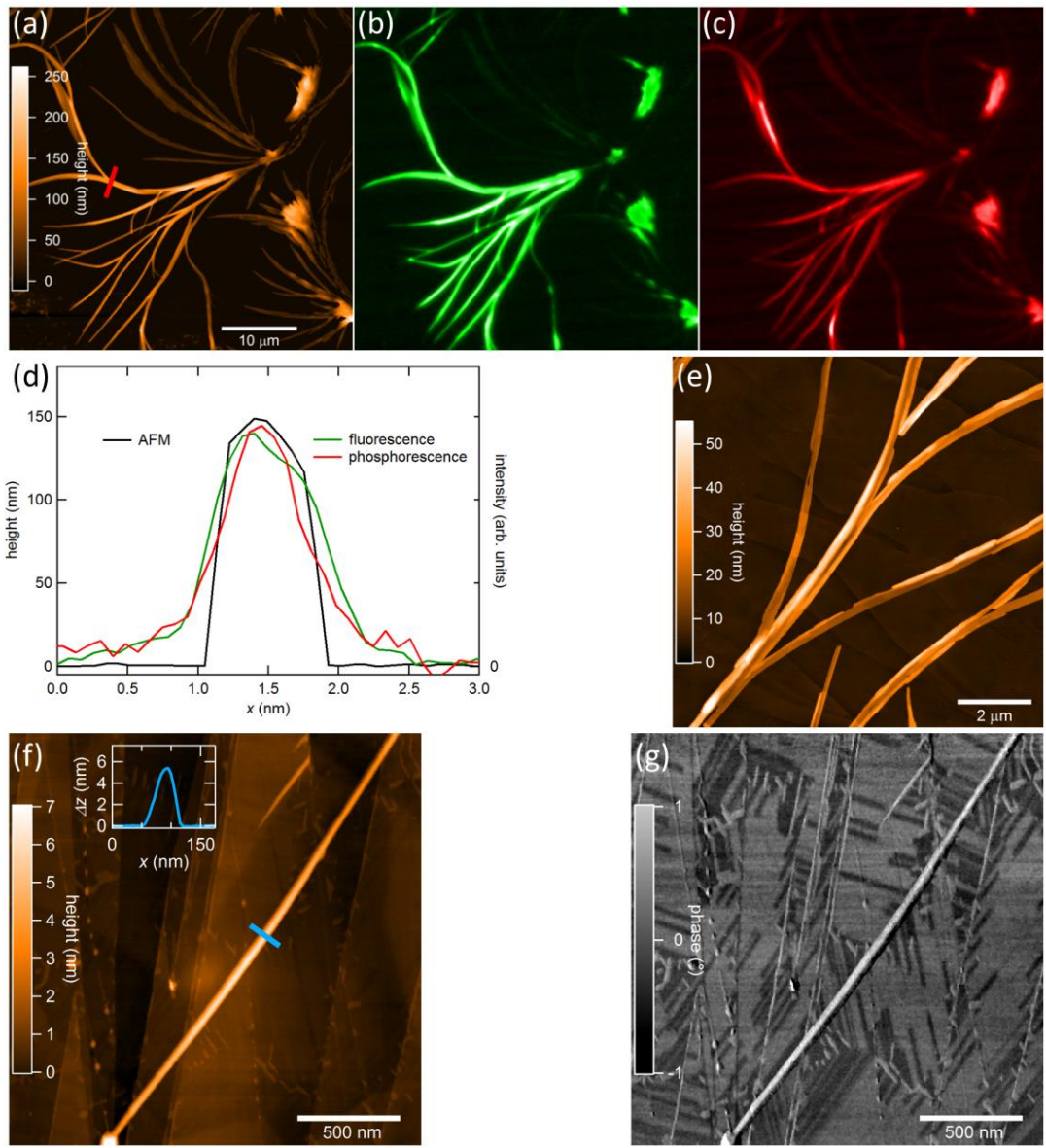

**Figure 2.** Geometry and luminescence of tree-like aggregates. (a) Survey AFM morphology, (b) correlative fluorescence image (exposure 20 seconds), and (c) correlative phosphorescence image (exposure 3 seconds). (d) Line sections along the path indicated in (a). (e) AFM example of a tree-like structure which seems to be composed of smaller, rod-shaped aggregates. (f) AFM topography and (g) phase of a sample after rinsing. Apart from a remaining thin strand (diagonal bright line, see inset for a line profile) domains of different polymorphs of monolayer aggregates are visible, most notably in the phase image (dark versus bright patches in (g)).

## 3.3 Deposition at modified evaporation rate

Spray deposition is an option to enhance solvent evaporation and may reveal whether crystallites already had precipitated prior to drop casting in the solution or not. Inspection by optical microscopy and AFM shows that the aggregates are much smaller and hardly exceed 40 nm heights; no tree-like aggregates have formed and predominantly isotropic, almost circular structures are found (figure 3(c)). So we conclude that the tree-like strands do not pre-form in solution, before contact with the surface, but rather in the course of the shrinkage of the droplet due to evaporation of the solvent. For spray deposition we use the same concentration like in drop casting and ordinary spray flasks made of glass and tubes of plastic. We applied one shot by a typical finger press from a large distance of about 75 cm. Smaller distances such as a few cm lead again to tree-like aggregates, probably because the deposited aerosol droplet diameter becomes already macroscopic or due to droplet coalescence. Another approach in structure variation – this time with lowered evaporation rate - is the aggregation under higher vapor pressure which was achieved by placing the samples inside a box during evaporation. This yields lengthy, curly, thin structures (figure 3 (d)).

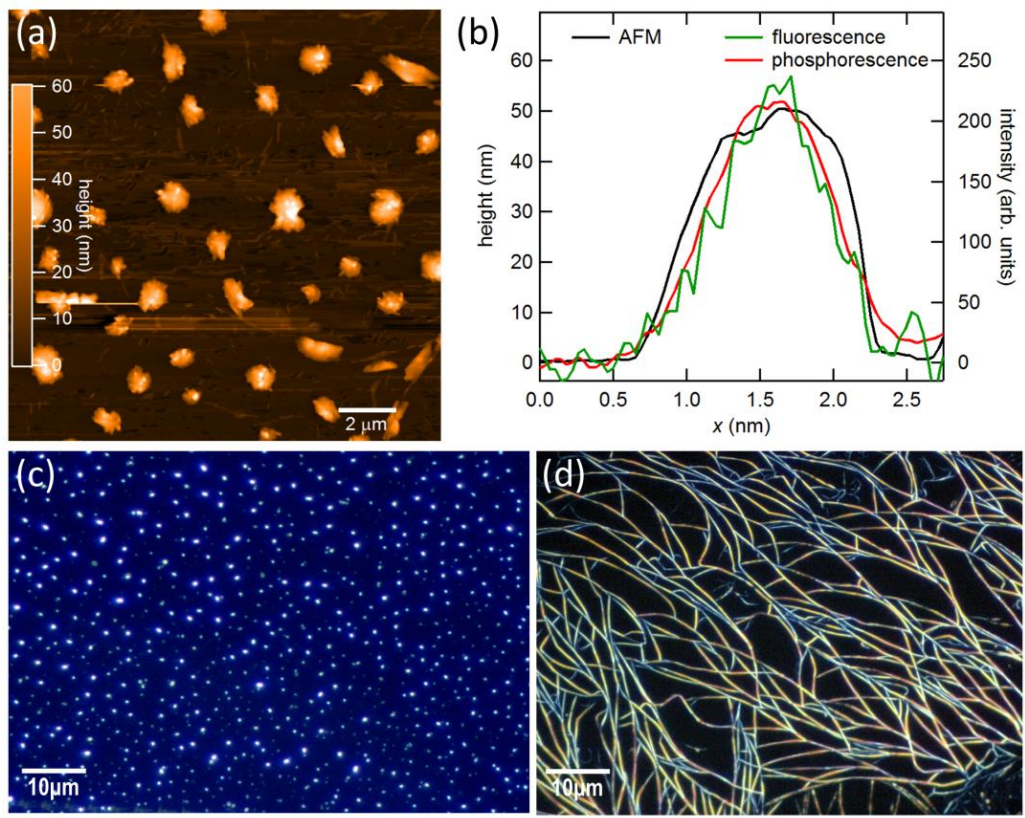

**Figure 3.** (a) AFM topography of almost isotropic aggregates. (b) Normalized line profiles from correlative AFM (black), fluorescence (green), and phosphorescence (red) micrographs along the path indicated in (a). (c) Optical dark field image of structures formed by spray deposition. (d) Dark field image of branched aggregates grown at increased vapor pressure.

*3.4 Luminescence intensity versus height*

We observe fluorescence in the region of the Q bands around 600 nm using a filter cube designed for the fluorescence marker "Texas Red". The emission window (595 nm - 665 nm) covers a substantial part of the fluorescence regime of Cu porphyrin. For this filter cube the excitation window (540 nm - 580 nm) does not include the Soret band (~400 nm); consequently relatively weak fluorescence may be expected.

However, a modified filter cube optimized for Soret excitation did not result in significantly higher signal. Due to this weak fluorescence the minimum height of objects with detectable emission beyond the noise floor is around 30 nm. Furthermore, we notice that the fluorescence regime is "contaminated" by Raman emission lines to a degree of up to 30 %. We observe strong phosphorescence in the transition energy regime of triplet states (800 nm - 1000 nm, see figure 4 (b)). Here, a long pass filter cube (>750 nm) is used to account for the spectral width of the signal, and the excitation window was 400 nm to 700 nm. Due to the different widths of excitation and emission windows used in the two filter cubes, the relative intensities of phosphorescence versus fluorescence are difficult to estimate on the basis of the microscopy data. Optical spectroscopy at 532nm laser excitation reveals that phosphorescence is considerably stronger than fluorescence, particularly after correcting for Raman emission (figures 4 (b)). In case of the monomer fluorescence is absent and phosphorescence dominates the emission. When comparing the luminescence of the different morphologies discussed above subtle differences in the intensity distribution of structures of similar height are observed, which we address in the following.

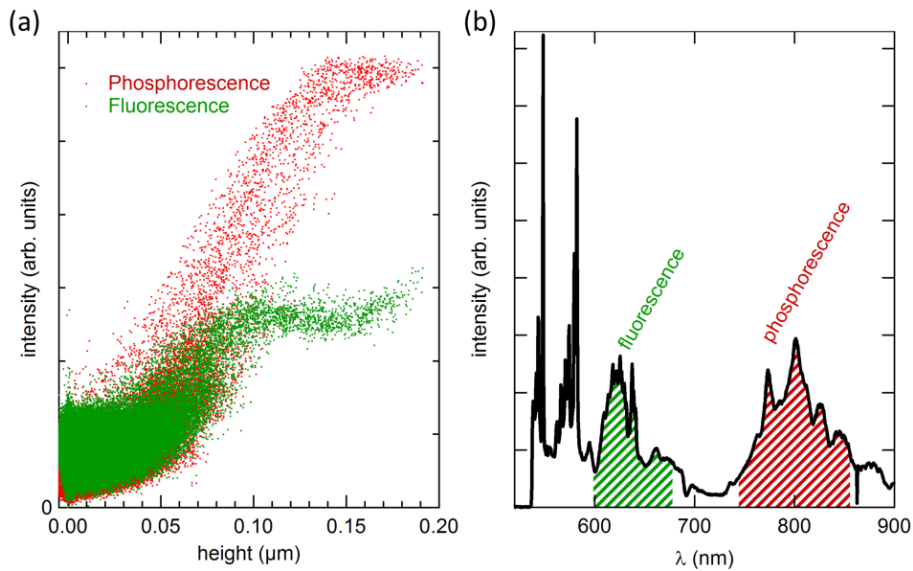

**Figure 4.** (a) Comparison of fluorescence and phosphorescence intensities of tree-shaped structures depending on aggregate height obtained by AFM. (b) Optical emission spectra of porphyrin aggregate strands upon laser excitation at 532nm. The sharp peaks correspond to Raman lines. These data were acquired in a confocal optical setup.

From correlated luminescence and AFM measurements we obtained the luminescence intensity as a function of structure height for a large number of objects (figure 4(a)). In the height regime of flat structures (below ~50 nm) the fluorescence intensity tends to rise parabolically with structure height; at intermediate heights the intensity rises about linearly with structure height; beyond a certain threshold height which is ~100 nm in case of fluorescence and ~130 nm in case of phosphorescence, the intensity appears to saturate, or even starts to decrease again as suggested by the fluorescence data (green data points in figure 4(a)). In the linear regime we evaluate the slopes of the behavior of intensity with respect to height [15]. To eliminate effects from the different illumination conditions and exposure times the comparison is applied to intensity slope ratios fluorescence/phosphorescence for the tree and isotropic structures respectively. This reveals higher ratios of the lengthy structures than for the isotropic ones. Hence, less non-radiative channels appear to be available for the lengthy structures, or the dipole transition probability for fluorescence is enhanced there. Possibly, upon assembly of the isotropic structures a higher degree of disorder is retained; the dewetting line moves substantially faster during aggregation of these structures and molecule transport is then less directional compared to the assembly of the lengthy, tree-like structures. Lower degree of order in the isotropic structures is further supported by roughness inspection. The lengthy structures exhibit AFM corrugations below around 1 nm while the isotropic structures appear 10 times rougher and thereby may contain more quenching sites. Also the ragged morphology of the isotropic aggregates suggests a lower degree of order or smaller domain sizes in compared to the strands. These observations corroborate the notion that the molecular ordering in the strands is better.

Sets of correlated microscopy images can be further inspected in order to address lateral effects. In particular cross sections can elucidate in how far luminescence does depend on morphological structure features. For the strands as well as the isotropic structures we observe that cross sections taken in AFM topographies are narrower than respective cross sections in luminescence images (figure 2c and 3c). This can be attributed to the lower spatial resolution of optical imaging versus AFM. In AFM the "resolution" is roughly proportional to the height, in optical microscopy the resolution a fixed value (at one wavelength). Luminescence appears to be rather homogeneously distributed over the structure and no hot spots or dark regions have been observed.

*3.5 Internal molecular ordering*

To learn about the orientation of the transition dipole moments in the strands polarization microscopy is employed. We illuminated by un-polarized light and acquired fluorescence maps with systematically varying polarization filter directions. The evaluation reveals an oscillatory behavior of fluorescence as a function of polarizer angle with angular period of 180°. The phase of these oscillations corresponds to the dominant polarization direction of the fluorescence.

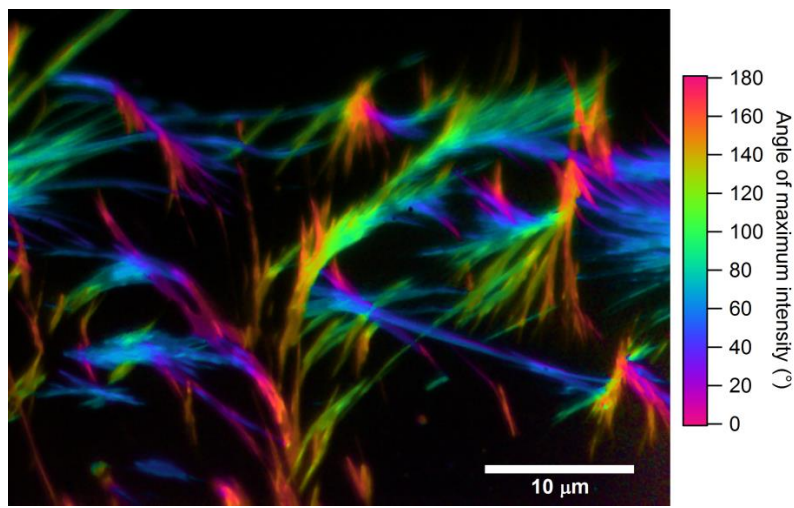

**Figure 5.** Dichroism map extracted from polarization-filtered fluorescence images under unpolarized excitation. The direction of maximum luminescence intensity is color-coded with respect to an arbitrary reference. Emission is predominantly polarized along the direction parallel to the strands. The observed fluorescence dichroism amounts to up to about 60% in this case.

Strands oriented along the same directions appear at similar colors in figure 5, i.e., the observed polarization direction closely follows the growth direction of the Cu porphyrin strands. Since our filter cube blocks the wavelength regime of the Soret band, we can only conclude on the transition dipoles involved in S1 and not on the S2 band. The observed fluorescence dichroism is up to 60%. The analysis points towards transition dipole moments oriented parallel to the strand axes. Due to the 4-fold symmetry of the monomer the observed period of 180° is merely compatible with internal structures where the plane of the individual molecules is oriented to some extent perpendicular to the substrate plane, i.e. "standing up". A continuation of the monolayers with flat-lying molecules as observed in [11] should have led to a 90° rotational period. The data are compatible with π-π structures such as brickstone-like, parallel-displaced, or herringbone arrangements. However localized stacking geometries do not account for the rather complex hierarchical structural organizations of porphyrin, such as helices and tubes [16–18]. In a PEEM study we observed pronounced Davydov splitting, as manifested by its dependence on excitation energy and polarization [19]. Davydov splitting is due to exchange interaction between non-translationally equivalent molecules in exited states [20], pointing towards a more complex unit cell.

Fluorescence lifetime imaging on the tree-like structures reveals lifetimes in the regime of a couple of nanoseconds. The spatial lifetime distribution of these structures is largely homogeneous. Also triplet-states showed quite little lifetime heterogeneity [21].

## 4. Discussion

*4.1 Optical density of states effects on excitation and emission*

Luminescence yields can become altered at the stage of absorption, at internal processes, and at emission. One may discriminate effects of the variation of the optical local density of states: vertically, i.e. due to mirroring at opaque substrates, or laterally, i.e. due to light landscape. The molecule structures may guide light, depending on their shape and size [22] which could lead to better or worse match between directions of the electric field and of transition dipole moments. Lengthy structures are obviously more likely acting as optical fibers and Poynting vector directions lie then preferably along the fiber axis. This could change absorption if the transition dipole moments of the aggregate are also oriented predominantly transversally with respect the fiber axis. With the isotropic shapes this would rather not result in modulated absorption. In view of absence of true resonator geometries, variations in substrate reflectivity or any metallic nanostructures, and due to insufficient information of the orientation of the molecules in the aggregates we focus our further discussion on internal processes and subsequent emission. A selection of possibilities is depicted in figure 6, this includes pathways associated with monomers [23] as well as options specific for oligomers and aggregates.

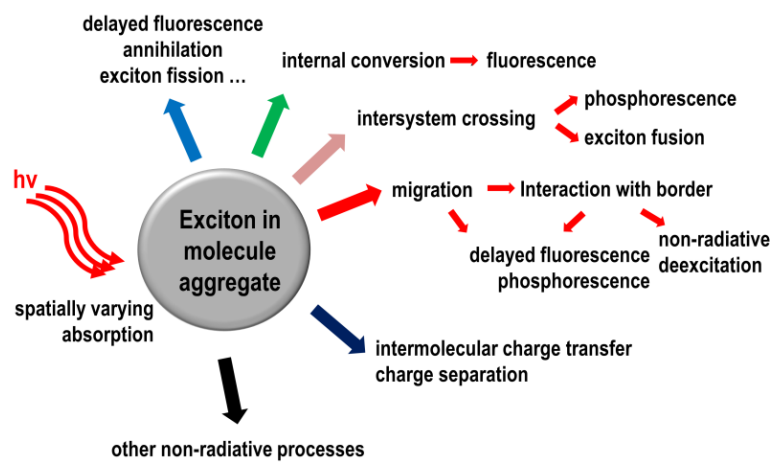

**Figure 6.** Overview of possible processes for de-excitation of electronic excitations in molecule aggregates.

*4.2 Aggregation effects on luminescence*

Most luminophores show weaker luminescence if aggregated compared to monomers. This is referred to as aggregation-caused quenching (ACQ) and is due to additional non-radiative de-excitation possibilities in the solid state. However a number of species show the reverse phenomenon, i.e. aggregation-induced emission enhancement (AIE) [24]. In fact, a different extent of order may modulate the degree of frustration of vibrational and rotational motion and thereby shift the relative ratio of non-radiative versus emissive de-excitation pathways. The observation of higher fluorescence yield versus phosphorescence of the structures with lengthy morphology as described above might originate from a kind of AIE. In particular the alkane side groups with their methyl ends attached to the meso-positions of the porphyrin could get locked in the dense aggregate and no longer be available for non-radiative energy transfer.

*4.3 Multi-reflection*

Which effect is responsible for the observed saturation, in both fluorescence and phosphorescence? The attenuation of incoming light rises with thickness of the aggregate, due to absorption and scattering effects. Aggregate thicknesses beyond the attenuation length will not exhibit further rise in intensity. Here the primary wavelengths are operative due to their generally lower mean free paths. Concerning emission internal reflection will give rise to multi-reflection at the dye-air and at the dye-silicon interfaces. The variation of refraction indices should then be moderate in the relevant optical wavelength regime. For porphyrin molecule aggregates we assume n to be nearby 2 [25], for silicon n is nearby 4. According to the Fresnel equations a node is expected at the interface between substrate and dye, while an antinode is expected at the interface between dye and air. To elucidate the effect of internal reflection of luminescence we performed simulations based on the Fresnel equations similar as in [26] using optical properties of related molecules [25]. The vertical variation of excitation intensity is modeled by the finite attenuation length inside the structure which is of the order of a few 100 nm. Luminescent light is treated to be monochromic with a wavelength of 630 nm for fluorescence, and 800 nm for phosphorescence, respectively. The observed emission intensities are calculated from an integral over the aggregate thickness where both absorption within the material and reflections at the interfaces are taken into account. Besides a smooth saturation the curves (not shown) exhibit an oscillatory behavior at larger heights which is also tentatively seen in figure 4(a) (green data points).

In order to verify the participation of internal multi-reflections and absorption giving rise to the observed curve shape of correlation plots such as shown in figure 4(a) we employ a similar analysis based on photoemission electron microscopy (PEEM). In contrast to earlier PEEM experiments where interference effects in have been utilized for obtaining object heights either using Lloyd's mirror geometry [27] or x-ray standing waves [28] we here evaluate optical interferences within the dye structure. The sample is illuminated by a pulsed laser ($\lambda = 269$ nm, see [21] and [19] for details) and the intensity of photo-emitted electrons is analyzed as a function of the local structure height. The advantage of PEEM is that it is only sensitive on a thin layer (<30 nm) close to the surface such that the integration over the entire aggregate thickness hardly smears out interference effects. Figure 7 shows the experimental data together with a fit to a simple model similar to the one described above. In contrast to optical microscopy only the incident light is relevant (which is monochromatic here) because the luminescence does not result in considerable photoelectron contribution. Pronounced oscillations are clearly visible and the good agreement with the model confirms the occurrence of optical internal reflections within the strands. The observation also implies that there are regions of PEEM intensity with an inverted dependence on the height which can result in an anticorrelation between PEEM and AFM images which we indeed observe for structures within a certain height region (figure 7(b)). This also means that PEEM intensities of 3D molecule structures are more influenced by the vertical optical density of states during excitation due to interference effects than luminescence.

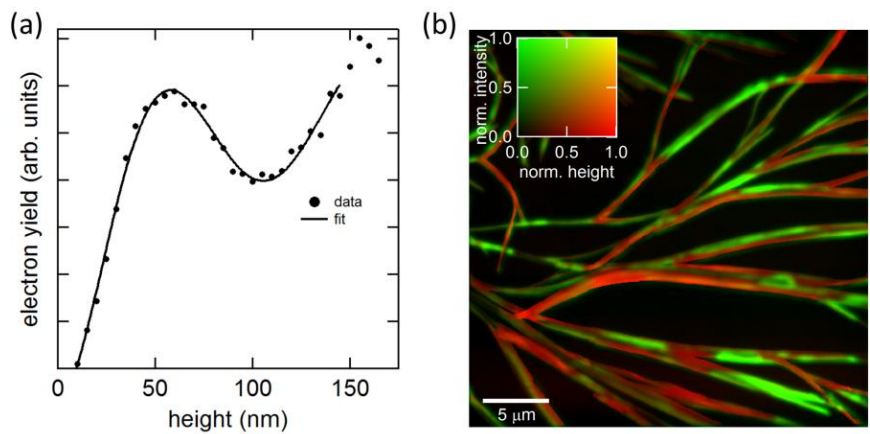

**Figure 7.** (a) Behavior of the PEEM photoelectron yield (laser excitation at 269 nm) as a function of height of the structure as measured by AFM. Dots: measurement; solid curve: fit to a multireflection model. (b) Correlative image of AFM height and photoelectron yield where red (green) areas correspond to large (small) heights and low (high) PEEM intensity. The brightness corresponds to the total intensity (PEEM) or height (AFM). The inset shows the color coding as a function of normalized AFM height and PEEM intensity. The occurrence of only red and green areas reveals an anticorrelation.

*4.4 Internal structure*

In principal different internal structure motifs inside the aggregates can be responsible for modulation of phosphorescence versus fluorescence ratios. ACQ and AIE are referring to variation of fluorescence yield versus molecule density. Another issue is changes of emission wavelengths. In case of one-dimensional crystallites red-shifted fluorescence transition is expected with J-type molecule stacking and blue-shifted with H-type [29,30]. In three-dimensional crystallites J- and H- type stackings are present at the same time, and the respective stacking arrangement will depend on the crystallographic axes considered. The elucidation of the internal structure is however not straightforward, because conventional diffraction methods require larger crystallites and more material. Another option would be high-resolution scanning probe microscopy such as in-liquid monitoring by AFM during dissolution or formation of the aggregates; solvent composition is then delicate because adhesion of the 3D structures on the surface must be sustained in the course of the process and during measuring.

**5. Conclusion**

Cu porphyrin aggregates have been prepared on surfaces such as silicon, glass, and HOPG and studied with respect to luminescence and morphology. The residence time of the dewetting front determines the aggregate morphology. Two types of morphologies have been observed: i) anisotropic, lengthy, tree-like shapes and ii) almost isotropic structures. Phosphorescence is generally favored over fluorescence and the local structure volume appears to be the main parameter governing luminescence yield. Lengthy tree-like structures do exhibit maximal fluorescence in proportion to phosphorescence compared to other morphologies such as flat or isotropic aggregates, and to monomers which exhibit an increasingly larger phosphorescence fraction. This is compatible with the effect of aggregation induced fluorescence enhancement and/or phosphorescence quenching. Fluorescence dichroism up to 60% is observed for the lengthy structure, being in accordance with substantial molecular order. Correlations of height determined by AFM with the luminescence intensity for the same sample area reveal a rather complex dependence. A simple model taking into account multi-reflections within the aggregates along the vertical direction can explain the observed behavior, as well as the pronounced height-dependent oscillations observed in photoemission maps. The observed variations of the fluorescence versus phosphorescence intensity appear to reflect the degree of order in the aggregates. This study serves as a basis to investigate in a next step metal-assisted luminescence enhancement [31] and migration effects in configurations where the molecule strands are coupled to plasmonic nanostructures.

**Acknowledgement**

We are grateful to the Deutsche Forschungsgemeinschaft for financial support of this work within the SFB 652 "Strong correlation and collective effects in radiation fields: Coulomb systems, clusters and particles" and within "Komplexe Molekulare Systeme" within the German Federal and Länder governments program. J.A.A.W.E. acknowledges the Council for the Chemical Sciences of The Netherlands Organization for Scientific Research, an ERC starting grant (Grant NANOCAT-259064) and the Ministry of Education, Culture and Science (Gravity Program 024.001.035). We thank Christin Baudisch (Schröter), Rike Broer, Regina Lange, Julia Heller, Tamam Bohamud, Lukas Rathje, and Mirjam Marsch for their help with data acquisition and analyses.